\documentclass[11pt,twoside]{article}

\usepackage{asp2004}
\usepackage{epsf}
\usepackage{psfig}
\usepackage{lscape}

\begin{document}
\title{Initial conditions for reaching the critical velocity}   %%% Fill in title
\author{Georges Meynet, Sylvia Ekstr\"om, Andr\'e Maeder, Fabio Barblan}   %%% Fill in author names
\affil{Geneva Observatory, CH-1290 Sauverny, Switzerland}    %%% Fill in author affiliations

\begin{abstract} %%% Abstract to run on from here.
The aim of this paper is to determine
the initial rotational velocities
required on the ZAMS for single stars
to reach the critical velocity, sometimes called the break-up velocity, during the Main-Sequence (MS) phase.
Some useful relations between $\Omega/\Omega_{\rm crit}$, $\upsilon/\upsilon_{\rm crit}$ ($\upsilon$ is the
velocity at the equator),
the moments of inertia, the angular momenta, the kinetic energy in the rotation and various
other basic physical quantities are obtained.
\end{abstract}

%%% MAIN BODY OF TEXT GOES HERE. CONSULT "INSTRUCTIONS FOR AUTHORS USING
%%% LATEX2E MARKUP", SECTIONS 2.3-2.6 FOR HELP WITH EQUATIONS, FIGURES,
%%% AND TABLES.

%\section{}   %%% Top level section head (remove "%" symbol)
%\subsection{}   %%% Second level section head (remove "%" symbol)
%\subsubsection{}   %%% Lowest level section head (remove "%" symbol)
%\section*{}	%%% Unnumbered top level section head (remove "%" symbol)
%\subsection*{}   %%% Unnumbered second level section head (remove "%" symbol)

\section{Stellar evolution near the critical limit}

The new results from VLTI on the shape of Achernar \citep{Do05} show that detailed parameters of 
rotating stars become accessible to interferometric observations. In order to make valuable comparisons with
models of rotating stars, we provide here a few basic data on rotating stellar models.

We also search the initial conditions required for a star to reach the critical limit during its
Main-Sequence phase. When the surface velocity
of the star reaches the critical velocity ({\it i.e.} the velocity such that the centrifugal acceleration exactly balances
gravity), one expects that equatorial ejections of matters
ensue \citep{Town04}. The physics involved in the ejection process, determining the quantity of mass ejected, the timescales between outburst episodes,
the conditions required for a star to present such outbursts are important questions not only for understanding
Be, B[e] or Luminous Blue Variable stars, but probably also for having a better knowledge on how the first
stellar generations in the Universe behaved \citep{EkTar}. Indeed, at lower metallicity, the stellar winds are weaker and
much less angular momentum is removed at the surface. This favors an evolution to the critical limit
and may have important consequences for the evolution of the first stellar generations. Also there are some indications that
the distribution of the initial velocities contains more rapid rotators at lower metallicity \citep{MG99}.
Finally, realistic simulations of the formation of the first stars in the Universe show that
the problem of the dissipation of the angular momentum is more severe at very low $Z$ than at the solar $Z$.
Thus these stars might begin their evolution with a higher amount of angular momentum \citep{Ab02}.

Rotation affects all the outputs of the stellar models. The reader will find reviews on the effects
of rotation in \citet{MMAA}, \citet{Hel00}, \citet{Ta04} and \citet{MT}.
In this work, we compute 112 different stellar models with initial masses equal to 3, 9, 20 and 60 M$_\odot$, for metallicities
$Z$ equal to 0, 0.00001, 0.002 and 0.020, and for values of the ratio of the angular velocity
$\Omega$ to the critical angular velocity $\Omega_{\rm crit}$ equal to 0.1, 0.3, 0.5, 0.7, 0.8, 0.9, 0.99. 
For the 3, 9 and 20 M$_\odot$ models, the computation was performed until the end
of the Main-Sequence phase or until the star reaches the critical velocity. For the 60 M$_\odot$ stars, we have obtained
all the models on the ZAMS, but only a subset of them were computed further on the Main-Sequence.
We consider that a star arrives on the ZAMS, when a fraction of 0.003 in mass of hydrogen has been burned
at the center. On the ZAMS, the star is supposed to have a solid body rotation. During the Main-Sequence phase the variation of $\Omega$ inside the star is computed self-consistently taking into account the various transport mechanisms, {\it i.e.}  convection, shear diffusion, meridional circulation and horizontal turbulence.
The removal of angular momentum at the surface by the stellar winds and the changes of $\Omega$ resulting
from movements of contraction or expansion are also accounted for. 
A detailed description of the 
physical ingredients used in this grid of models will be given elsewhere (Ekstr\"om et al., in preparation), let us just mention
here that the prescriptions for the opacity tables, the initial chemical compositions and the treatment of convection are as in \citet{MMXI}.

\begin{figure}%[!t]
\plottwo{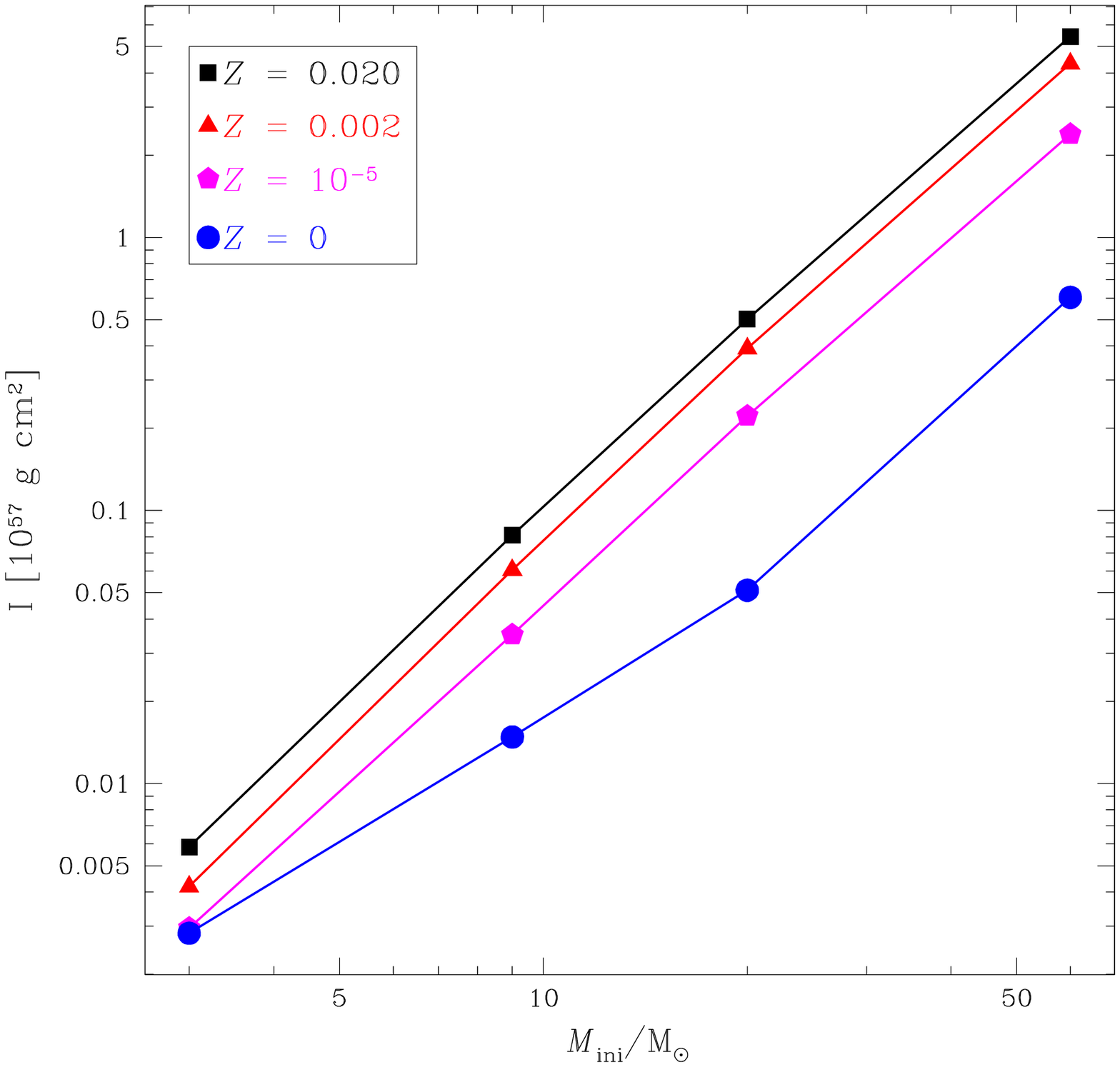}{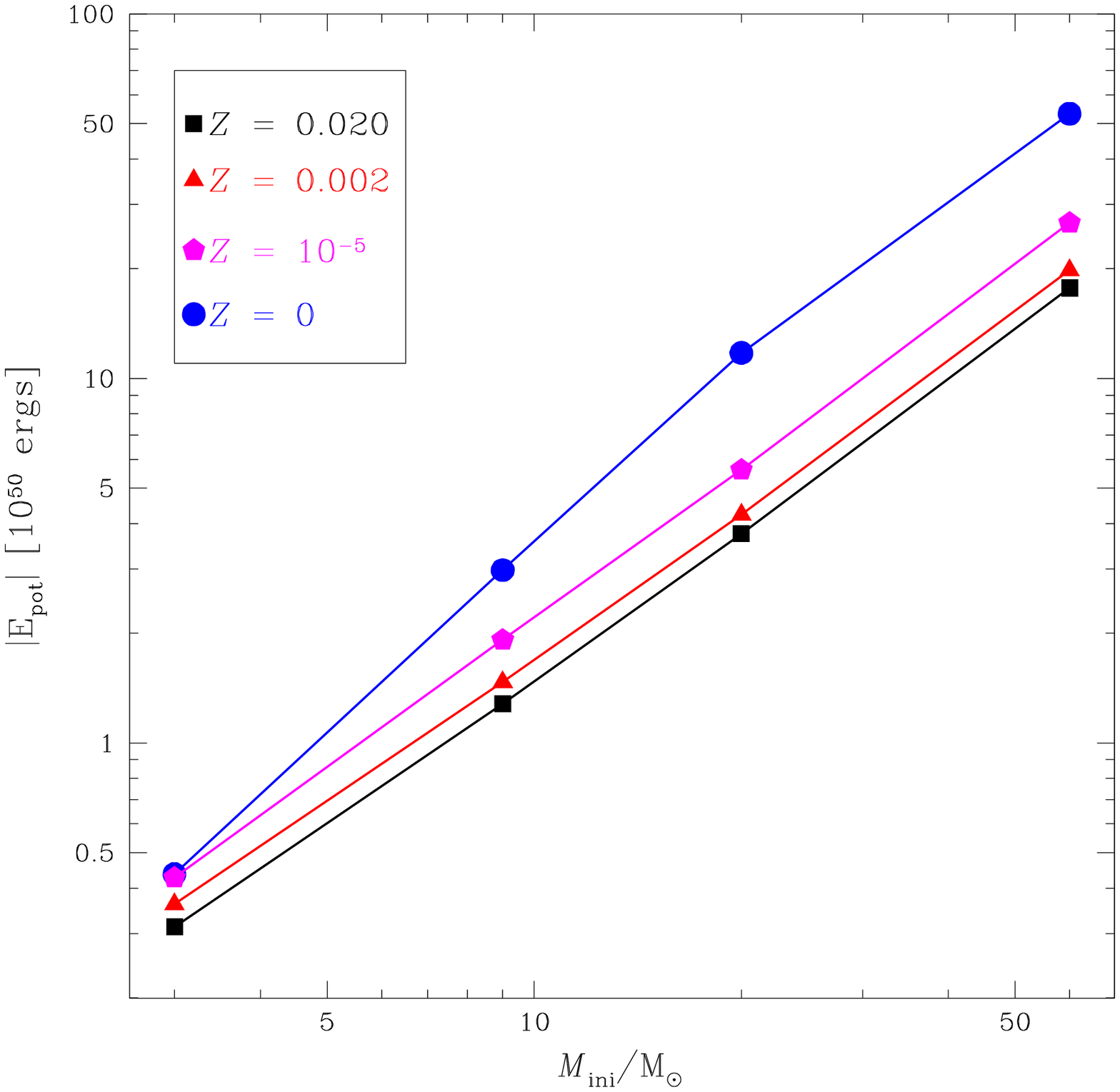}
\caption{{\it Left:} Variations of the momentum of inertia on the ZAMS as a function of
the initial masses for various metallicities $Z$ (squares: $Z=0.020$; triangles:
$Z=0.002$; pentagons: $Z=0.00001$ and circles: $Z=0$).
{\it Right:} Variations of the gravitational energy on the ZAMS as a function of
$M_{\rm ini}$ for various $Z$.
}
\label{moin}
\end{figure}

\begin{figure}%[!t]
\plottwo{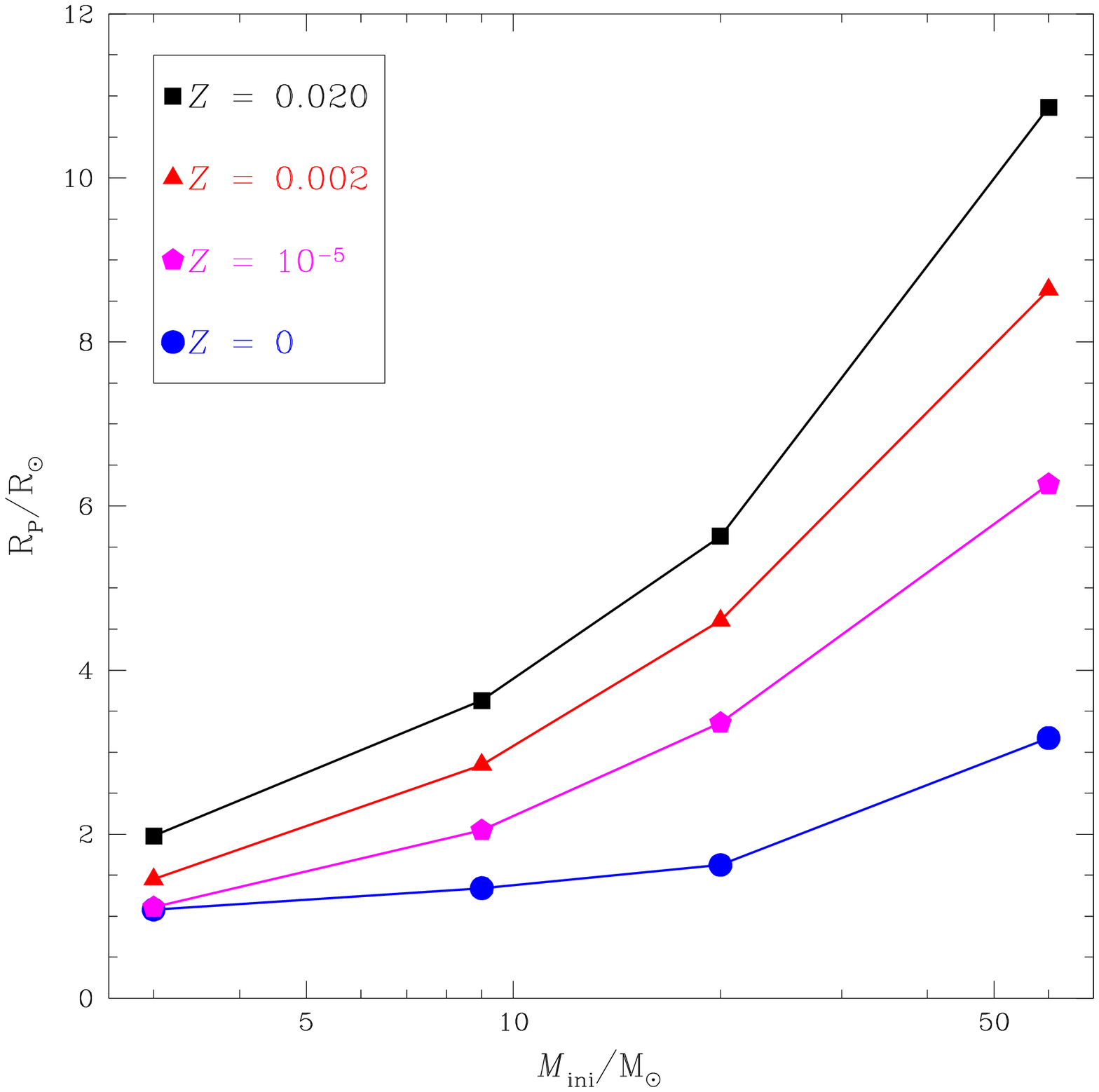}{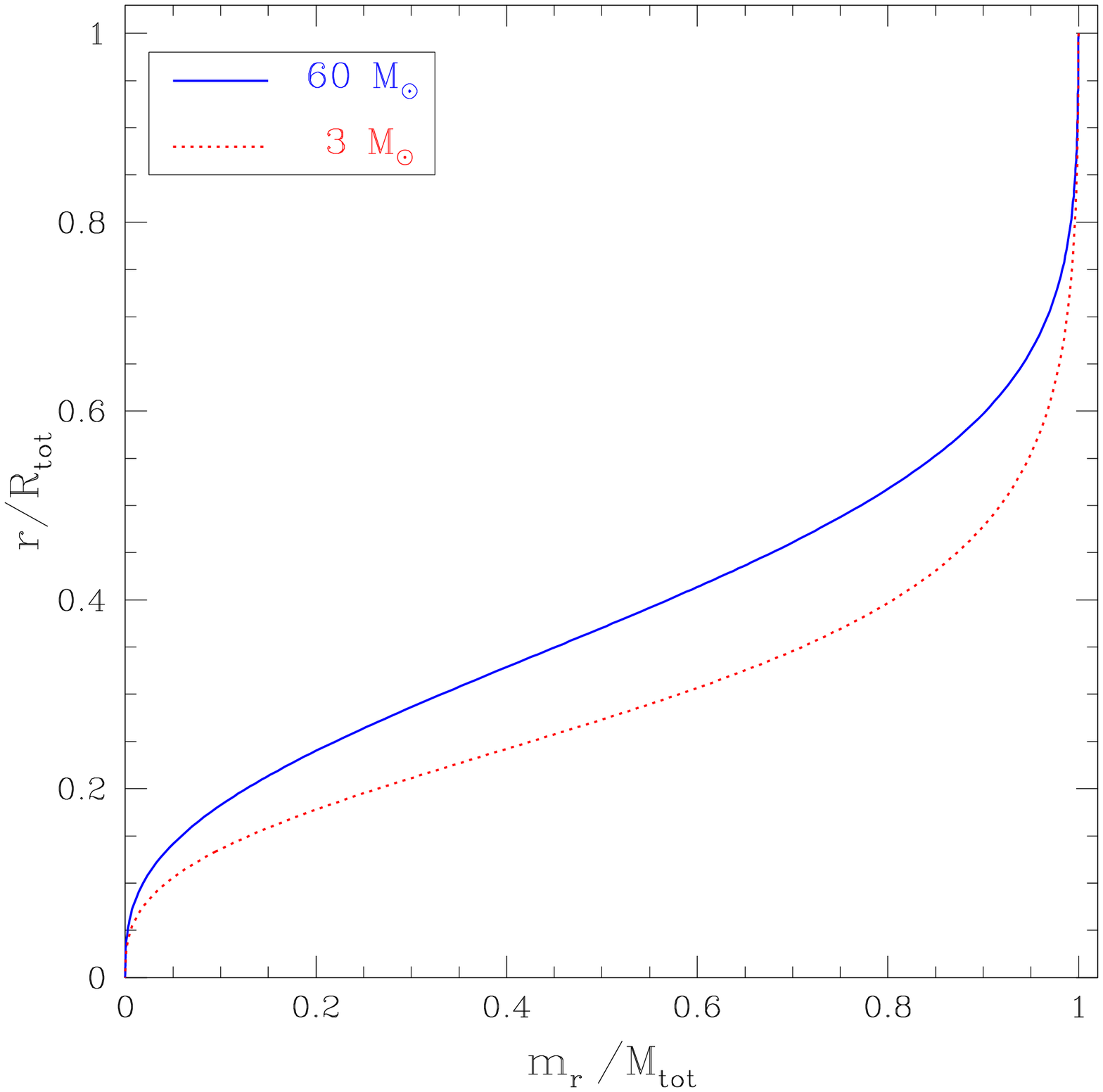}
\caption{{\it Left:} Variations of the polar radius on the ZAMS as a function of
$M_{\rm ini}$ for various metallicities. {\it Right:} Fraction of the total mass (m$_r$ is the
mass inside the radius $r$) as a function
of the fraction of the total radius for solar metallicity models with
$\Omega/\Omega_{\rm crit}$=0.1.
}
\label{radii}
\end{figure}

%\begin{figure}%[!t]
%\plottwo{eueg.eps}{eradeg.eps}
%\caption{{\it Left:} Ratio of the internal energy  to the gravitational energy
%as a function of $\Omega/\Omega_{\rm crit}$
%for various initial masses and metallicities. {\it Right:} Ratio of the energy
%in radiation to the gravitational energy as a function of $\Omega/\Omega_{\rm
%crit}$ for various initial masses and metallicities.
%}
%\label{energ}
%\end{figure}

\begin{figure}%[!t]
\plottwo{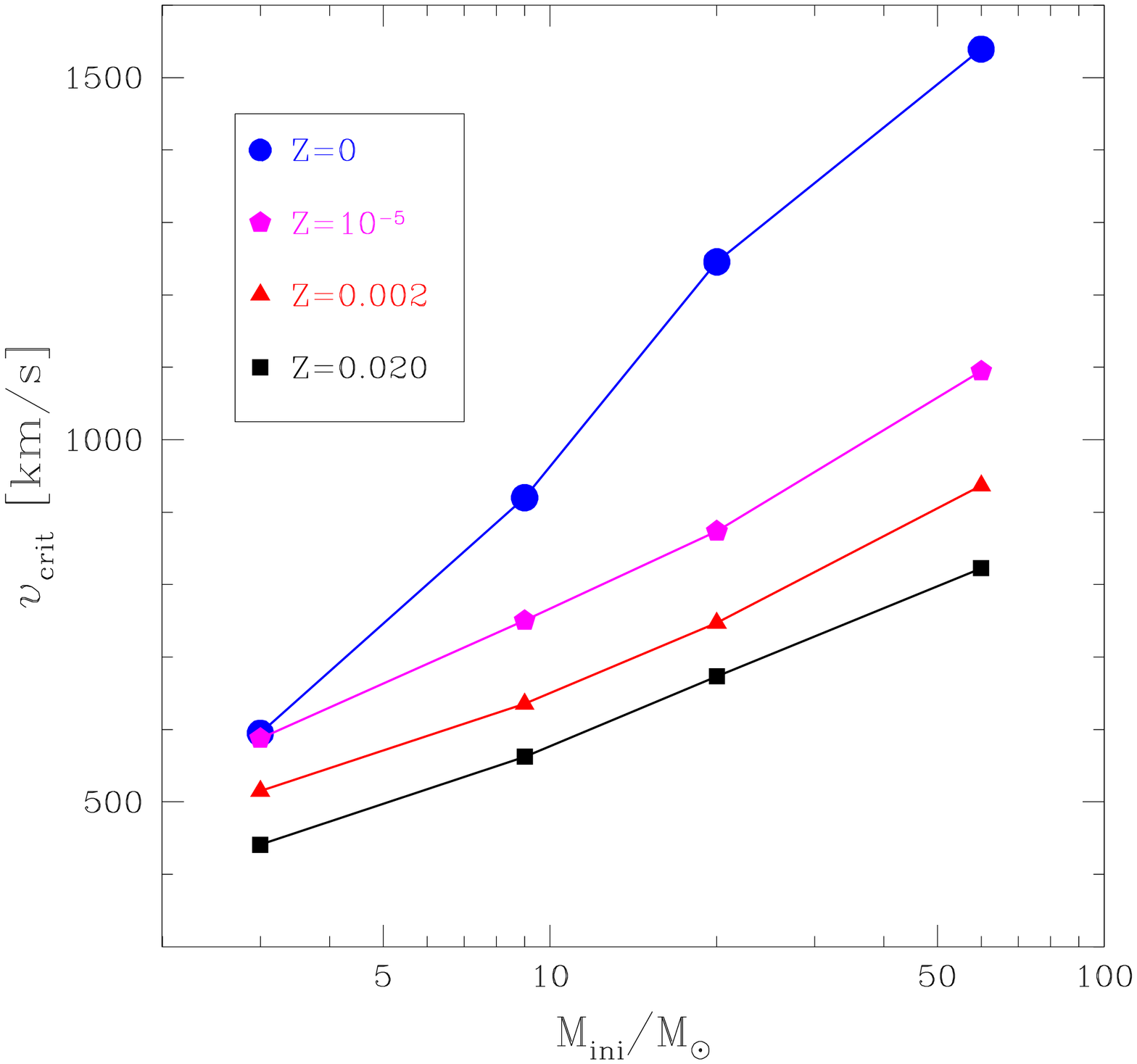}{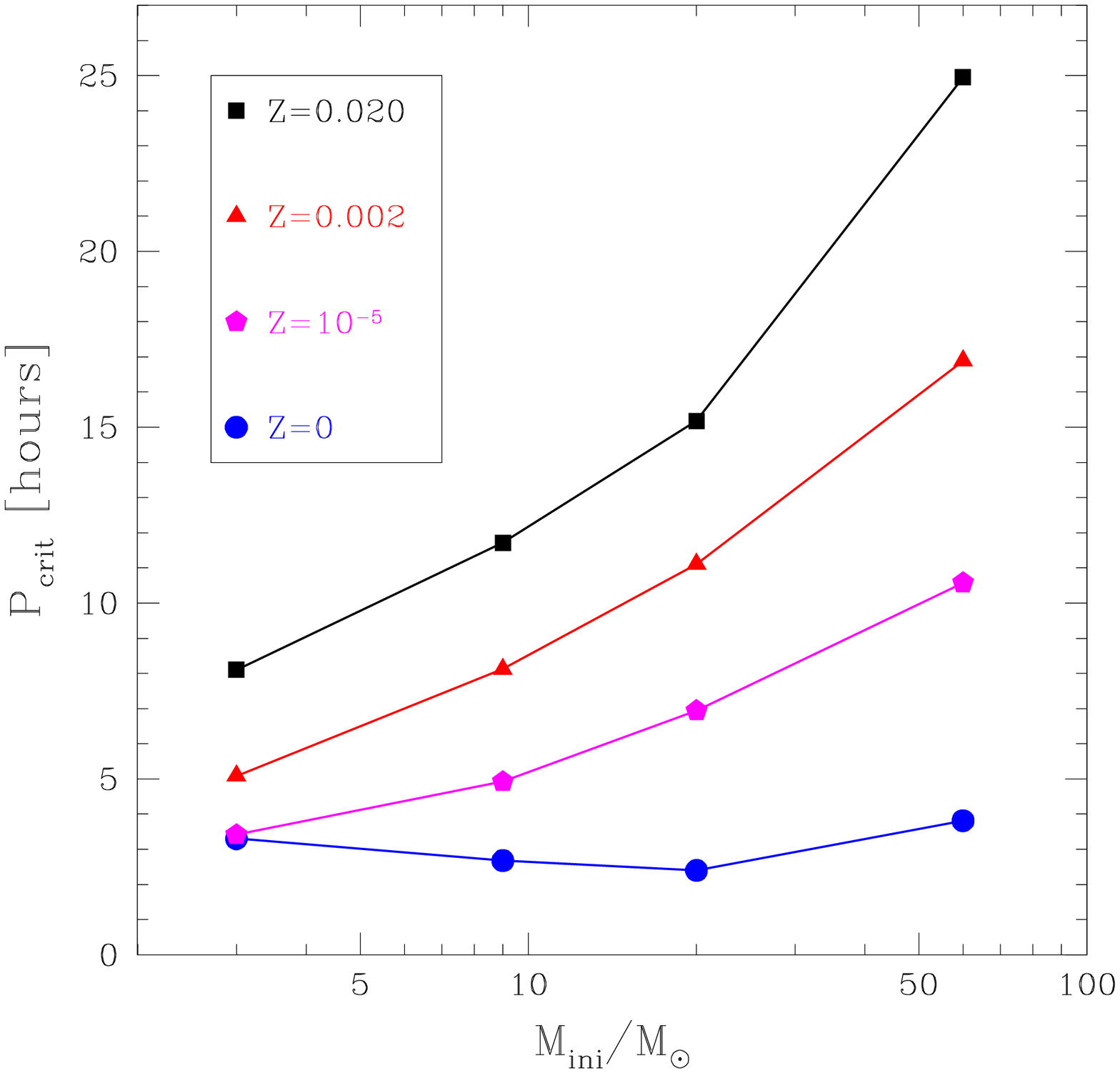}
\caption{{\it Left:} Variation of the critical equatorial velocity  on the ZAMS
as a function of the initial mass for various metallicities. {\it Right:}
Variation of the critical rotation period on the ZAMS as a function of the
initial mass for various metallicities.
}
\label{vcrit}
\end{figure}

Based on the set of numerical results described above,
we first present useful relations between
basic quantities, and then 
explore the initial conditions required for a star to reach the critical velocity during
it MS evolution.

\section{Moments of inertia and energy contents on the ZAMS}

\begin{figure}%[!t]
\plotone{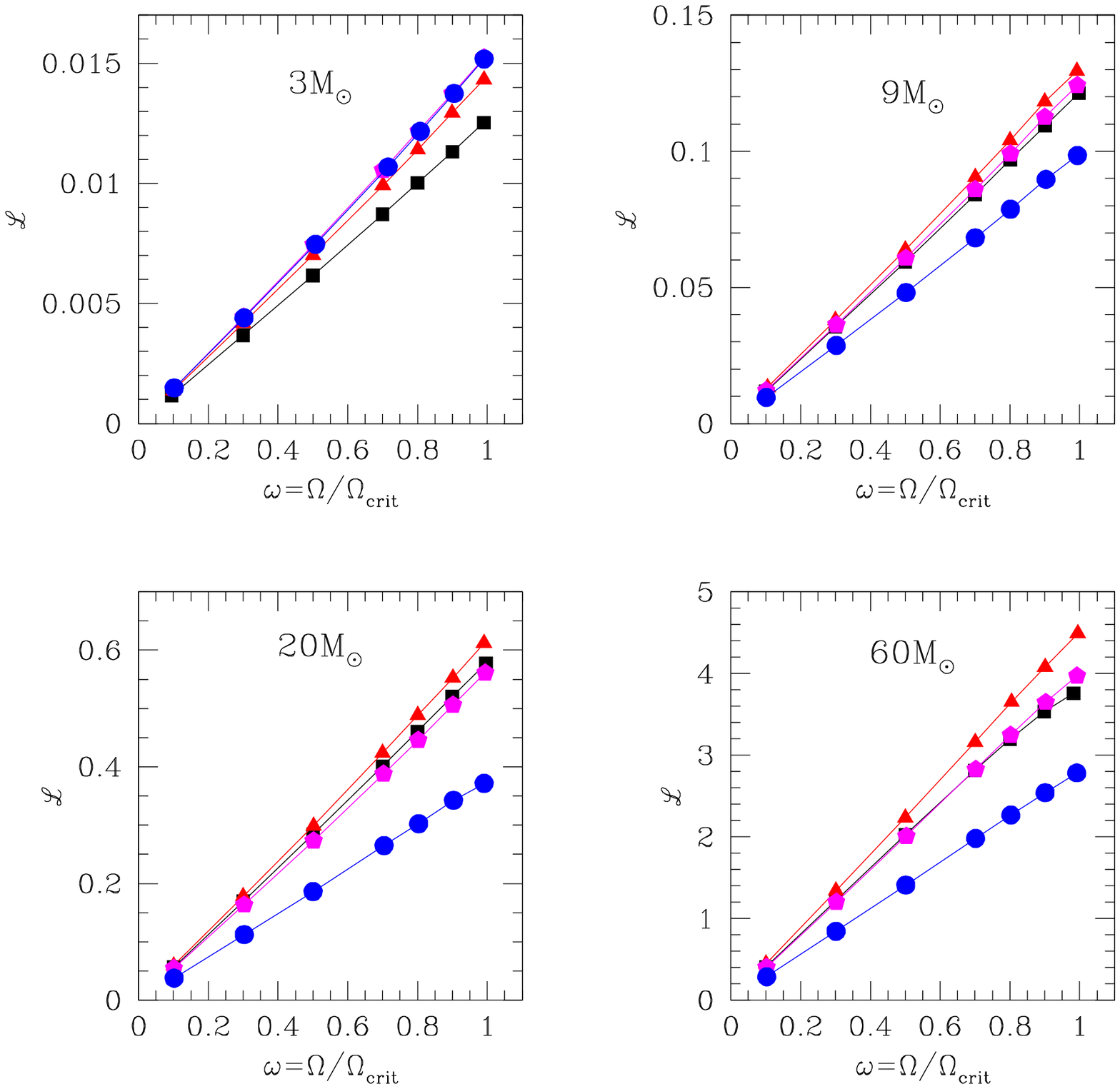}
\caption{Variations of the angular momentum on the ZAMS as a function of
$\Omega/\Omega_{\rm crit}$ for various metallicities, plotted according to the
initial mass of the models. Symbols as in Fig.~\ref{moin}. The angular momentum is
given in units of 10$^{53}$ g cm$^2$ s$^{-1}$. Same symbols as in Fig.~\ref{moin}.
}
\label{ltot}
\end{figure}

Figure~\ref{moin} shows for different initial masses and metallicities
the moment of inertia and the gravitational energy of the stellar models on the ZAMS.
Only the results obtained for a value of $\Omega/\Omega_{\rm crit}$ equal to 0.7 are presented, since
these quantities present a very weak dependence on $\Omega/\Omega_{\rm crit}$.
Numerically, these quantities were computed by summing the contribution of the different shells
composing the star (between 250 and 400 shells). Since the deformation due to rotation is very weak
in most of the star (even near the critical velocity) and since the outer layers contain little mass, we
neglect, in all the following estimates, the deformation of the shells induced by rotation. 
To each isobaric surface, limiting a volume $V$ (which is not a sphere), we associate an average radius $r$ given by $r=(V/(4/3)\pi)^{1/3}$.    

\begin{figure}%[!t]
\plotone{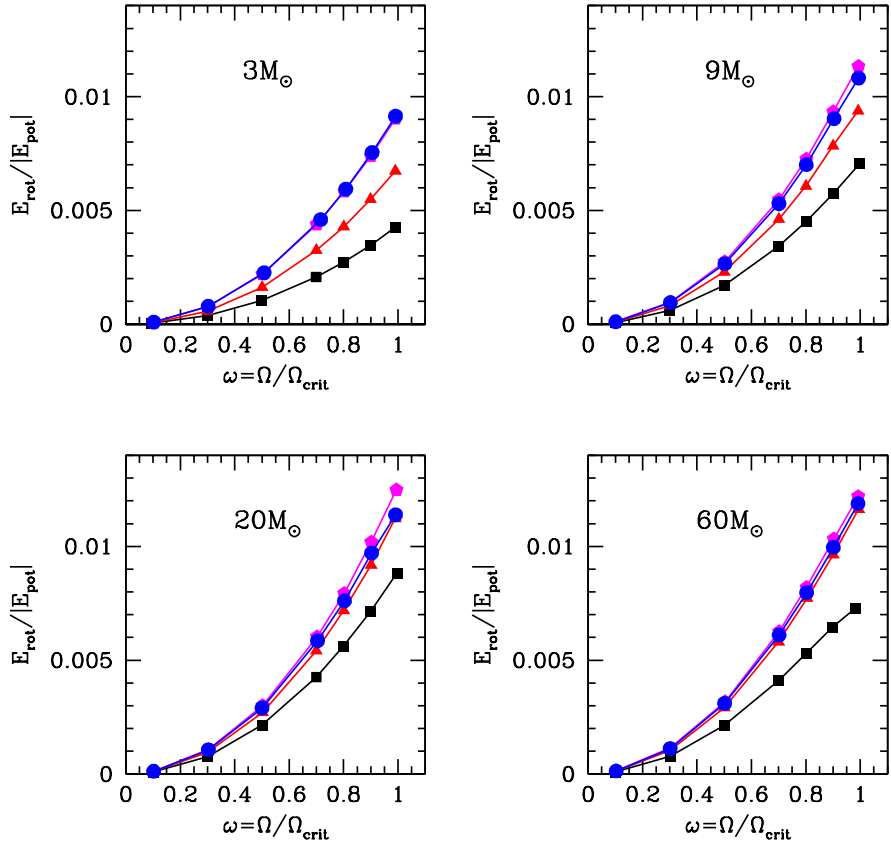}
\caption{Ratio of the kinetic energy in the rotation to the gravitational energy
as a function of $\Omega/\Omega_{\rm crit}$ for various metallicities, plotted
according to the initial mass of the models. Same symbols as in Fig.~\ref{moin}.
}
\label{ereg}
\end{figure}

Looking at the results concerning the moments of inertia, we note that, at a given metallicity, the moments of inertia increase with the initial mass.
The increase amounts to about 3 orders of magnitudes between 3
and 60 M$_\odot$. For the non-zero metallicities, 
we have linear relations between log I and log $M$.
In the case of the Pop III stellar models, the linear relation breaks down and
the increase of the moment of inertia with
the mass is less steep than at higher metallicities. 

What is the cause of this difference ?
Passing from 3 to 60 M$_\odot$ increases the radius by a factor 
5-6, whatever the metallicity between $Z=$ 0.00001 and 0.020 (see the left part of Fig.~\ref{radii}).
On the other hand, for zero metallicity stars, the radius passes from about 1 R$_\odot$ for a 3 M$_\odot$ stellar model to ~3.2 R$_\odot$ for a 60 M$_\odot$, thus the increase amounts to only a factor 3.
This comes from the fact that, at the beginning of the core H-burning phase
in Pop III massive stars, only the pp chains are active. The pp chains are not
efficient enough to compensate for the energy lost at the surface and the star
must extract energy from its gravitational reservoir.
The contraction is more severe in the higher mass range where the luminosities are higher.
This explains the shallower increase of the radius as a function of the mass in Pop III stars and
therefore the less steep increase of the moment of inertia with the mass.

Let us add here that the contraction occurring in Pop III stars at the beginning
of the core H-burning phase leads to an increase of the central temperature until
the  point of helium ignition is reached. Then,  
due to He-burning, small amounts of carbon
and oxygen are synthesized. When the abundances of carbon and oxygen in the core reach a level of 10$^{-10}$ in mass fraction, the H-burning can occur through the CNO cycle and this cycle
becomes, as at higher metallicity, the dominant energy source for the rest of the Main-Sequence phase.

Looking at the left part of Fig.~\ref{moin}, we note that, for a given initial mass, the variation as a function of the metallicity is much smaller than the variation as a function of the initial mass at a given metallicity. 
There is at most an order of magnitude of decrease when the metallicity passes from Z=0.020 to 0.
The decrease is more pronounced in the range of the high mass stars than in the range of the intermediate 
mass stars. The reason for that is again the difference in the way the beginning of the H-burning occurs
in massive Pop III stars. 

\begin{figure}%[!t]
\plotone{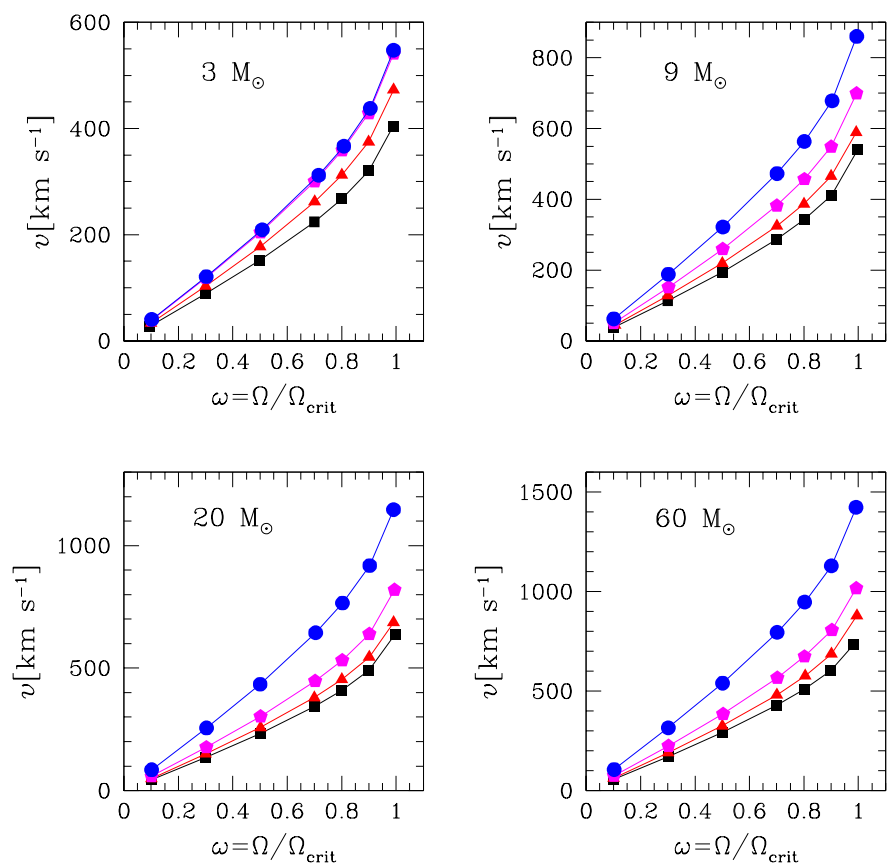}
\caption{Variations of the equatorial velocity at the surface on the ZAMS as a
function of $\Omega/\Omega_{\rm crit}$ for various metallicities, plotted
according to the initial mass of the models. Same symbols as in Fig.~\ref{moin}.
}
\label{vsurm}
\end{figure}

The gravitational energy or binding energy is shown in the right part of Fig.~\ref{moin}. 
Passing from 3 to 60 M$_\odot$, the binding energy increases by about two orders of magnitude.
For a given initial mass, the dependence on the metallicity is much more modest.
For a 20 M$_\odot$ stellar model at $Z=0.020$ the binding energy is about 4$\cdot$10$^{50}$ ergs,
its $Z=0$ counterparts has a binding energy equal to slightly more than 10$^{51}$ ergs, {\it i.e.}
about a factor 2.5 higher. 
Let us recall for comparison that
the binding energy of a neutron star is of the order of 10$^{53}$ ergs, about two orders of magnitude
greater.  

The left part of Fig.~\ref{radii} shows how the polar radii of the different models vary as a function
of the initial mass and metallicity. The right part of this figure illustrates the fact that
in the 3 M$_\odot$ stellar model, a given fraction of the total mass is enclosed in a smaller fraction of the total radius than in the 60 M$_\odot$, indicating that lower mass stars have a more concentrated mass
distribution than higher mass stars. 

%\begin{figure}%[!t]
%\plotone{epot.eps}
%\caption{Variations of the gravitational energy on the ZAMS as a function of
%$M_{\rm ini}$ for various $Z$.
%}
%\label{epot}
%\end{figure}

%Figure~\ref{energ} (left part) shows the ratios of the kinetic energy in the gaz to the gravitational energy.
%The dependance on the initial rotation and on the metallicity is very weak. The initial mass
%is the key parameter. As expected the kinetic energy in the gaz is a smaller fraction of the gravitational energy
%in the most massive stars. The most massive stars present a higher energy content in the radiation (see the right part
%of the figure). The sum of the kinetic energy in the gaz and of the energy in the radiation fields amounts to about half
%the gravitational energy, which is consistent with the Virial theorem.

%\begin{figure}%[!t]
%\plotone{vsurz.eps}
%\caption{Same as Fig.~\ref{vsurm} but with the different initial masses at the
%same metallicity represented in the same plot. Continuous line: 3 M$_\odot$;
%dotted: 9 M$_\odot$; dashed: 20 M$_\odot$; long-dashed: 60 M$_\odot$.
%}
%\label{vsurz}
%\end{figure}

\section{Critical velocity, angular momentum, kinetic energy of rotation and
surface velocity on the ZAMS}

\begin{figure}%[!t]
\plottwo{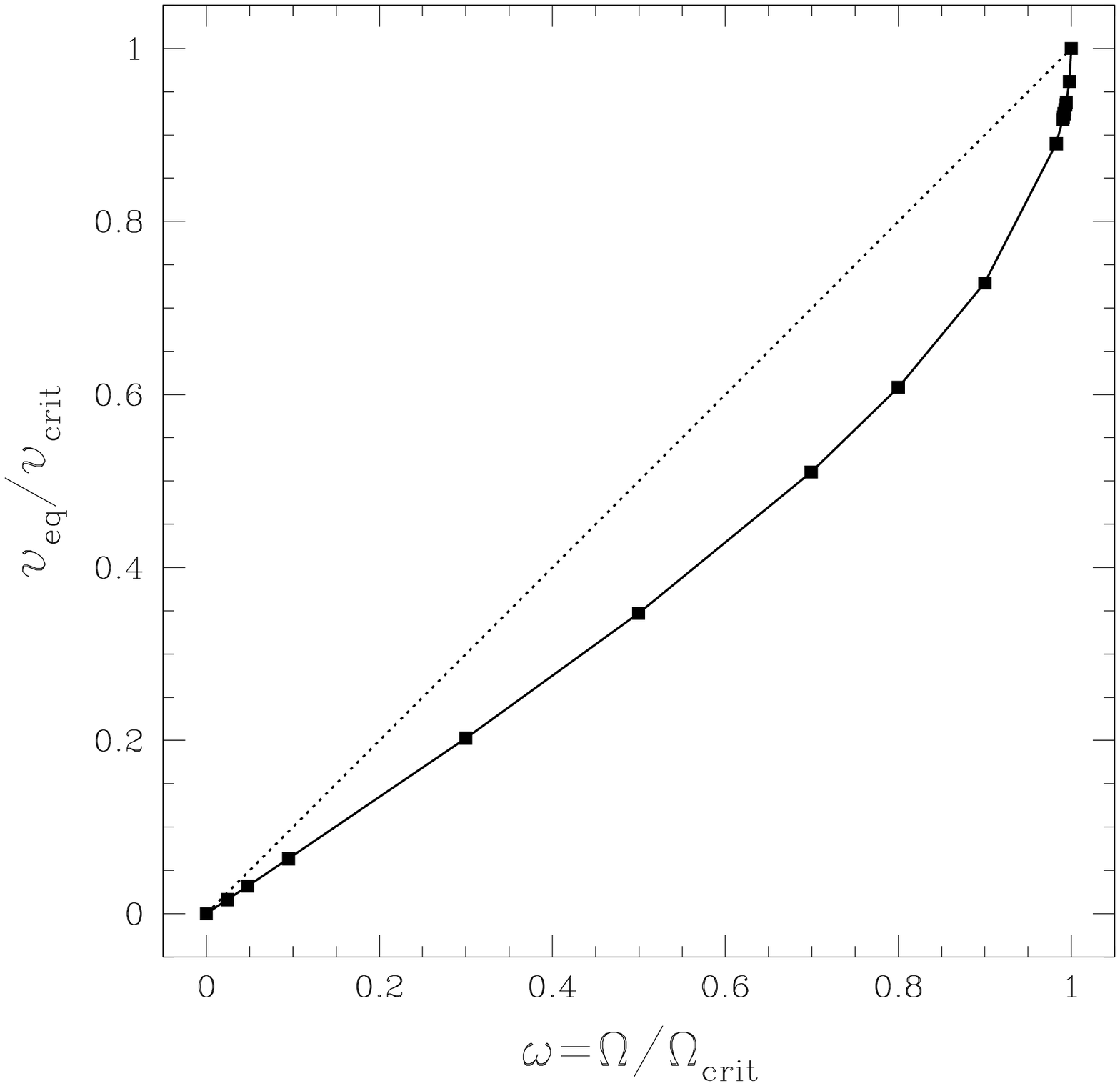}{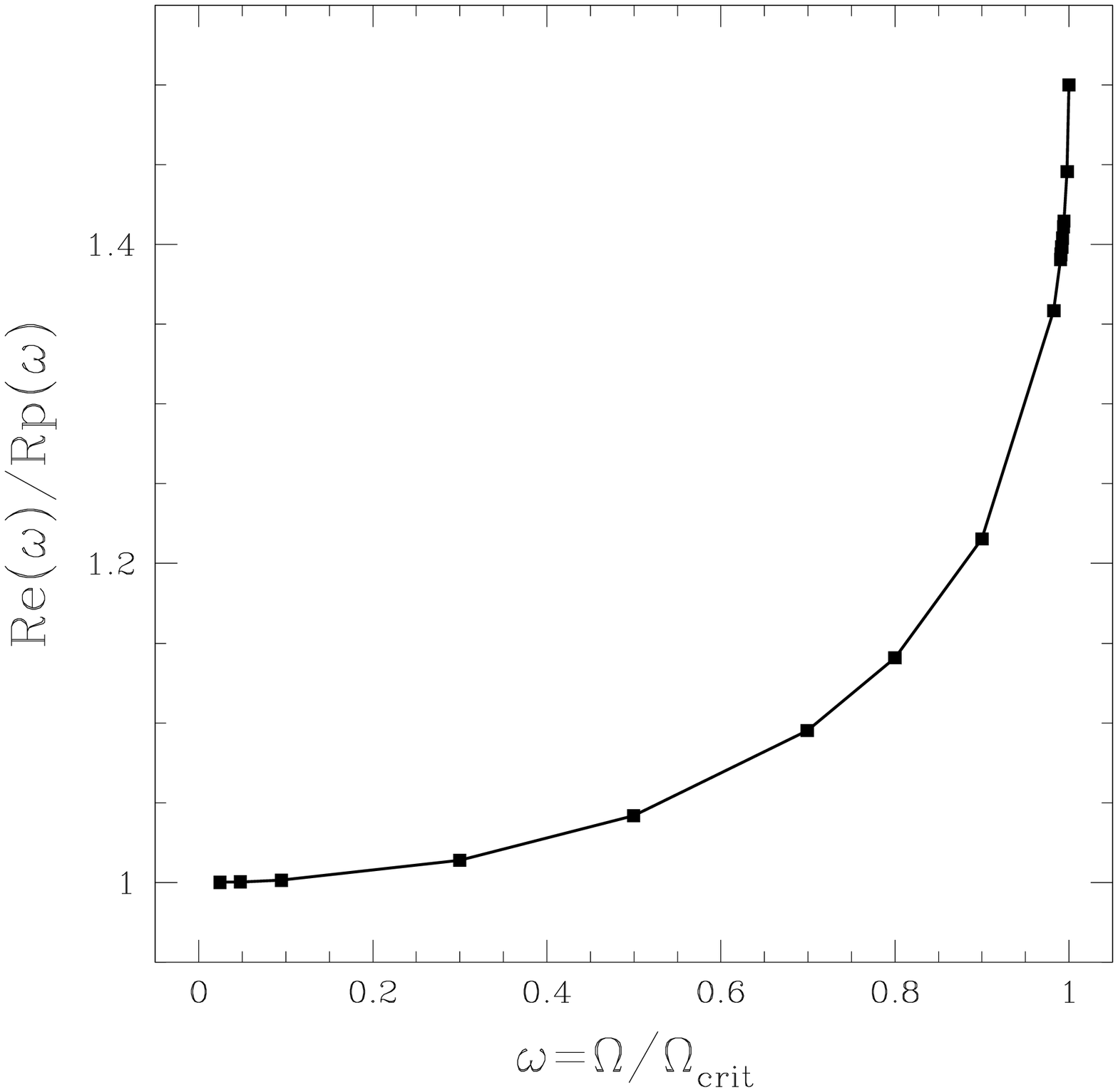}
\caption{{\it Left:} Variation of $\upsilon/\upsilon_{\rm crit}$ as a
function of $\omega=\Omega/\Omega_{\rm crit}$. {\it Right:} Variation of the
ratio $Re(\omega)/Rp(\omega)$ (equatorial over polar radius) as a function of
$\omega=\Omega/\Omega_{\rm crit}$. These relations have been obtained in the
frame of the Roche model. They do not depend on the mass, the metallicity or
the evolutionary stage considered.
}
\label{ovc}
\end{figure}

The variation of the critical velocity as a function of the initial mass and metallicity is shown 
in Fig.~\ref{vcrit}. At a given metallicity, the critical velocity is higher for higher initial mass stars.
The increase is  not far from being linear. The slope remains nearly the same for metallicities
higher than 0.00001, while it is much steeper for the pop III stellar models.
For a given initial mass, the critical velocity is higher at lower $Z$.
For the 60 M$_\odot$ model, the critical velocity passes from slightly more than 800 km s$^{-1}$ 
at Z=0.02 to nearly 1600 km s$^{-1}$ at $Z$=0.

The critical periods are shown on the right part of Fig.~\ref{vcrit}. They are related to
$\Omega_{\rm crit}$ by the simple relation $P_{\rm crit}= 2\pi/\Omega_{\rm crit}$.  
The critical
periods are comprised between 2.5 and 25 days for the whole domain of masses and metallicities explored
in the present work.

The variation with $\Omega/\Omega_{\rm crit}$ of the total angular momentum content is shown
in Fig.~\ref{ltot}. Note that the highest point on the right of the figures gives an estimate of what could be called the critical angular momentum content.
Since the moment of inertia of the star does not depend much on the rotation velocity,
and since on the ZAMS, we supposed solid body rotation,
one obtains linear relations between $\mathcal{L}$ and $\Omega/\Omega_{\rm crit}$. We note that the
metallicity dependence remains modest at least in the range of metallicities between 0.00001 and 0.020.
For the 9 M$_\odot$ models and above, starting from the Z=0.020 models, there is 
first a slight increase of the angular momentum content when lower metallicities are considered, the initial mass and the
value of $\Omega/\Omega_{\rm crit}$ being kept constant
(compare the curve with the black squares for the Z=0.020 models with the curve with the triangles
corresponding to 0.002 models). Then for still lower metallicities, the angular momentum decreases with the
metallicity (the curve for the Z=0.00001 models overlap the Z=0.02 models) reaching its lower values for the
Pop III stellar models.

%\begin{figure}%[!t]
%\plotone{rayons.eps}
%\caption{Variations of the polar radius on the ZAMS as a function of
%$M_{\rm ini}$ for various metallicities.
%}
%\label{rayons}
%\end{figure}

In the mass range between 9 and 20 M$_\odot$, 
two stars on the ZAMS, having the same angular momentum content, would have very similar
value of $\Omega/\Omega_{\rm crit}$ whatever their metallicity between 0.00001 and 0.02. Only if the
metallicity is zero, would the value of $\Omega/\Omega_{\rm crit}$ corresponding to that angular momentum content be 
much higher. As a numerical example, 
a 20 M$_\odot$ on the ZAMS with $\mathcal{L}$= 0.3$\cdot$10$^{53}$ cm$^2$ g sec$^{-1}$ has a
$\Omega/\Omega_{\rm crit}=~0.5$ when Z is comprised between 0.00001 and 0.020. This angular momentum
content corresponds to $\Omega/\Omega_{\rm crit}=~0.8$ when Z=0.

The rotational kinetic energy expressed as a fraction of the binding energy is shown in Fig.~\ref{ereg}.
Rotational energy amounts to at most a percent of the binding energy. This is consistent with
the well known fact, that the effects of the centrifugal acceleration remains quite
modest on the hydrostatic structure of the stellar interior, even when the surface rotates with a velocity near the critical one.

\begin{figure}%[!t]
\plottwo{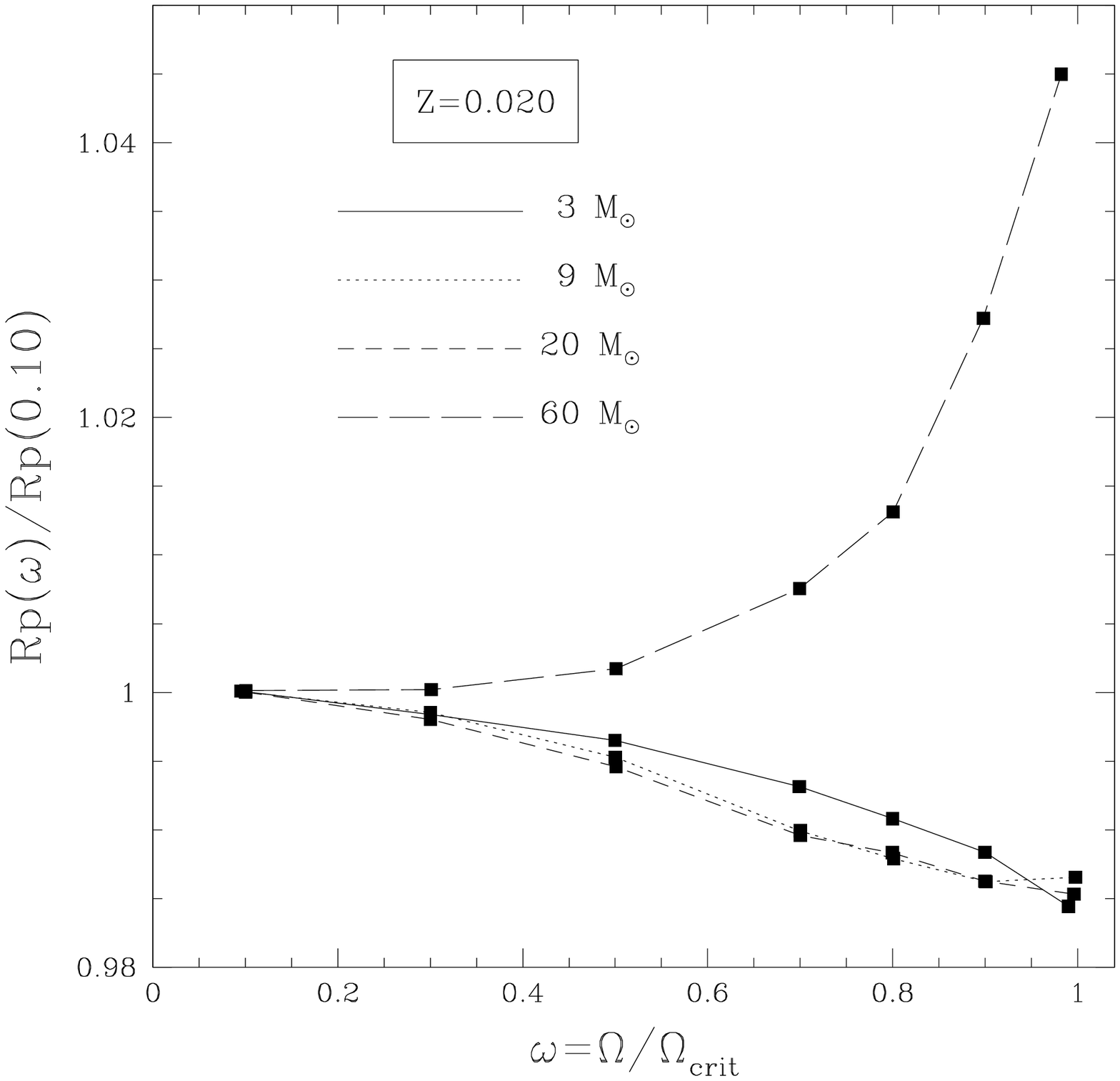}{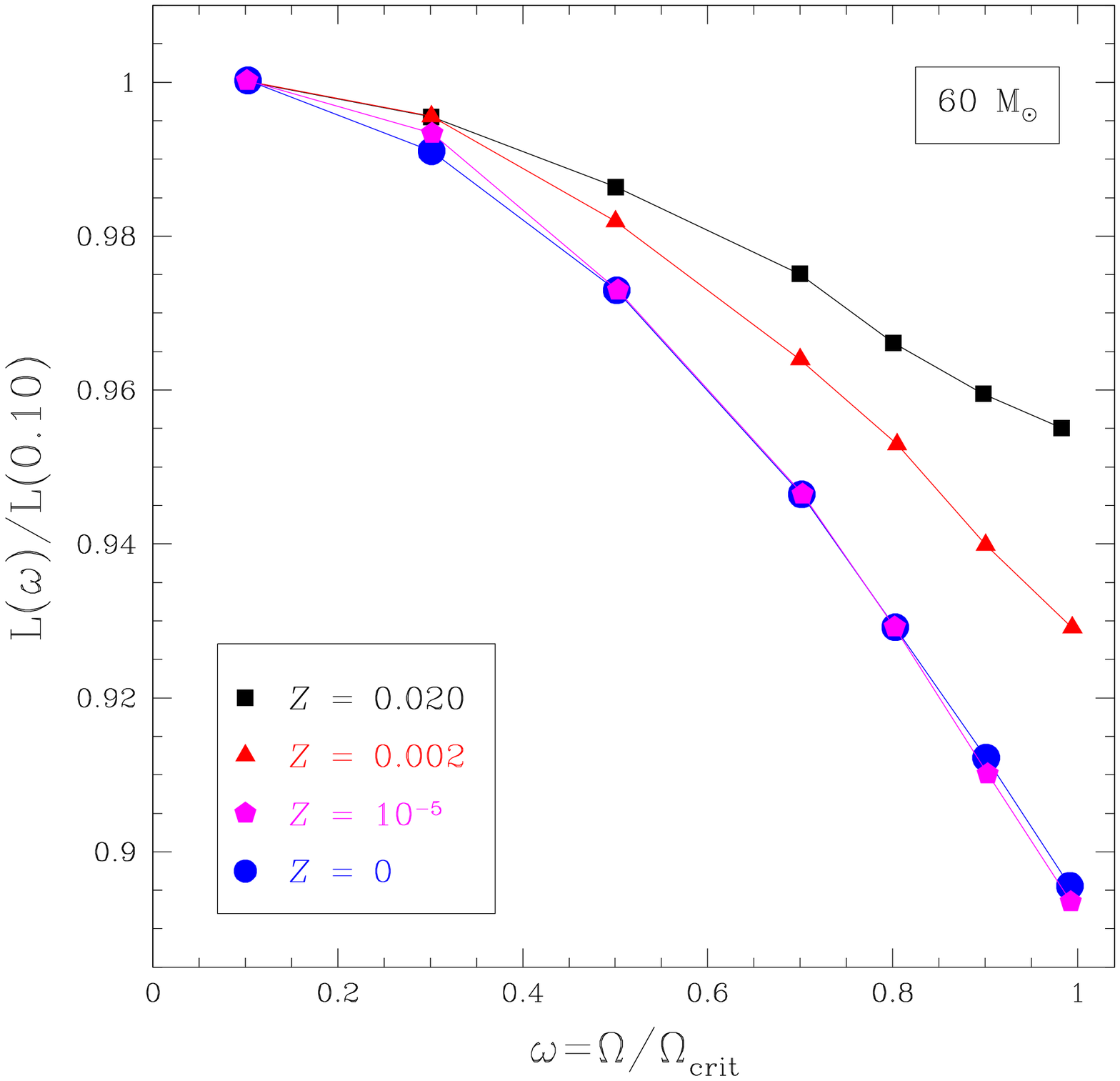}
\caption{{\it Left:} Variation of the polar radius $R_{\rm p}(\omega)$ as a
function of $\omega=\Omega/\Omega_{\rm crit}$, normalized to the value of
$R_{\rm p}$ at $\omega=0.10$, for various initial masses at $Z=0.020$. {\it
Right:} Variation of the total luminosity $L(\omega)$ as a
function of $\omega=\Omega/\Omega_{\rm crit}$,, normalized to the value of $L$
at $\omega=0.10$, for the 60 M$_\odot$ models at various metallicities (squares:
$Z=0.020$; triangles: $Z=0.002$; pentagons: $Z=10^{-5}$ and circles: $Z=0$). 
}
\label{rprpo}
\end{figure}

\begin{table}[t]
\caption{Velocity in km s$^{-1}$ on the ZAMS corresponding to $\Omega/\Omega_{\rm crit}=0.7$} \label{tbl-1}
\begin{center}\scriptsize
\begin{tabular}{ccccc}
M   &  $Z$=0              & $Z$=0.00001 & $Z$=0.002   & $Z$=0.020   \\
    &                     &             &             &             \\
\hline
    &                     &             &             &             \\
3   &     312             &     301     &     262     &     224     \\
9   &     473             &     383     &     325     &     287     \\
20  &     644             &     447     &     380     &     344     \\
60  &     795             &     566     &     480     &     428     \\                   
    &                     &             &             &             \\
\hline
\end{tabular}
\end{center}
\end{table}

The relations between the equatorial velocity and $\Omega/\Omega_{\rm crit}$ are plotted in
Fig.~\ref{vsurm}. We give in Table~\ref{tbl-1} the values of $\upsilon$ obtained for 
$\Omega/\Omega_{\rm crit}=0.7$ for various masses and metallicities. The value
$\Omega/\Omega_{\rm crit}=0.7$ corresponds to an initial rotation of ~290 km s$^{-1}$ for
a 9 M$_\odot$ stellar model. The time averaged velocity of this model during the Main-Sequence phase is equal to  $\sim$240 km s$^{-1}$.

We see that for a given value of $\Omega/\Omega_{\rm crit}$ and of the initial mass, the surface equatorial velocity is higher at lower $Z$. 
At a given metallicity, a  value of $\Omega/\Omega_{\rm crit}$ corresponds to
a value of the equatorial velocity which is higher for higher initial masses.
This means that if the stars begin their evolution on the ZAMS with 
approximately the same value of $\Omega/\Omega_{\rm crit}$, they have a higher
surface velocity at lower metallicities and for higher initial masses. For instance, 
a 3 M$_\odot$ stellar model with an initial value of $\Omega/\Omega_{\rm crit}$ equal to 0.7 has
an initial velocity equal to 220 km s$^{-1}$ (see Table~\ref{tbl-1}), while 
a 60 M$_\odot$ model, starting with the same value of $\Omega/\Omega_{\rm crit}$, has
an initial velocity of 800 km s$^{-1}$. 

%\begin{figure}%[!t]
%\plotone{qcc.eps}
%\caption{Variation of the mass fraction of the convective core on the ZAMS
%as a function of $\Omega/\Omega_{\rm crit}$
%for various metallicities, plotted according to the initial mass of the models.
%}
%\label{qcc}
%\end{figure}

\section{The Roche model and surface deformation on the ZAMS}

The relation between $\upsilon/\upsilon_{\rm crit}$ and
$\Omega/\Omega_{\rm crit}$ obtained in the frame of the Roche model
is shown in Fig.~\ref{ovc} (left part). This relation is the same whatever the initial mass
or metallicity. One sees that except at the two extremes, the values of
$\upsilon/\upsilon_{\rm crit}$ is smaller that that of
$\Omega/\Omega_{\rm crit}$. 
The variation with $\Omega/\Omega_{\rm crit}$ of the ratio 
of the equatorial radius to the polar radius is also a property of the
Roche model. It is shown in the right part of Fig.~\ref{ovc}.
We note that for $\Omega/\Omega_{\rm crit}$ below 0.7, the equatorial radius is longer than the polar one by less than 10\%.

%Figure~\ref{rayons} shows the polar radius in units of solar radius for different initial mass stars
%at various metallicities. The polar radius depend weakly on rotation. the data shown in Fig.~\ref{rayons}
%correspond to $\Omega/\Omega_{\rm crit}=0.7$.
%We note the very weak dependance of the radius as a function of the mass for Pop III stars.
%The variation with the mass increases when higher metallicities are considered. 

%\begin{figure}[!t]
%\plotfiddle{dhr9.eps}{10cm}{-90}{50}{50}{-200}{300}
%\caption{Evolutionary tracks in the HR diagram for the 3 M$_\odot$ stellar models.
%The first set of tracks on the left corresponds to Z=0 models, then are plotted from 
%right to left the models for respectively Z = 0.00001, 0.002 and 0.020. The initial rotation considered
%for each set are $\Omega/\Omega_{\rm crit}$=0.1, 0.3, 0.5, 0.7, 0.8, 0.9 and 0.99.}
%\end{figure}

\begin{figure}[!t]
\plotone{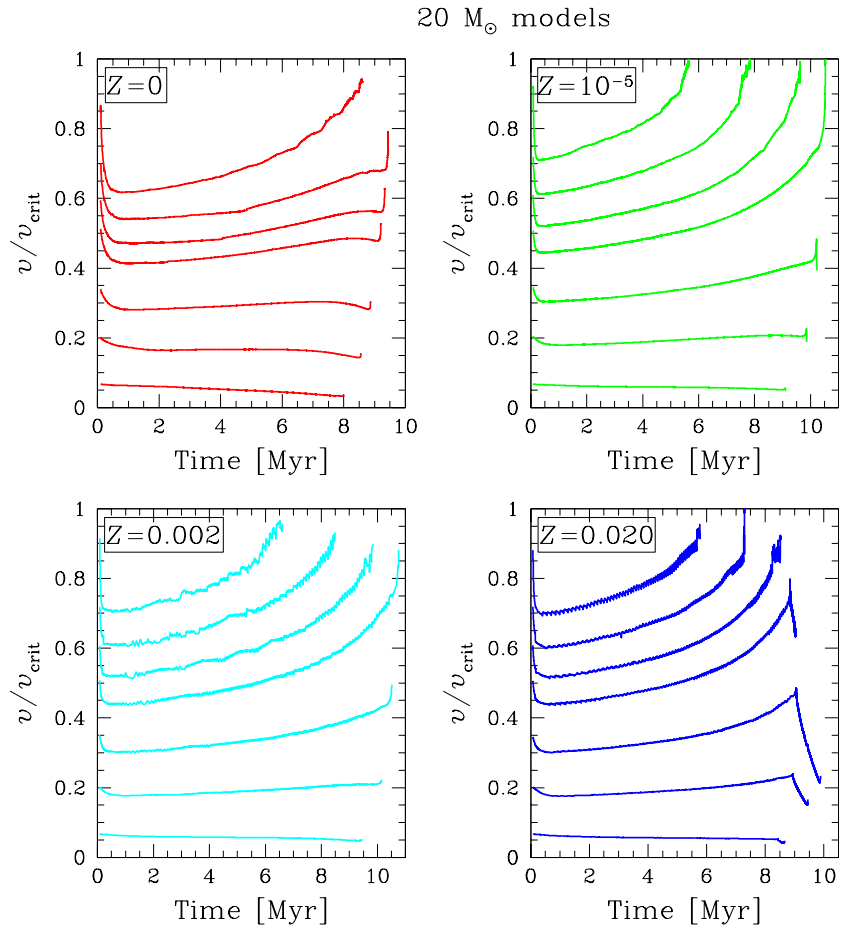}
\caption{Evolution as a function of time of $\upsilon/\upsilon_{\rm crit}$ at the surface
of 20 M$_\odot$ stellar models for different initial metallicities and velocities.From bottom
to top, the models with initial values of $\Omega/\Omega_{\rm crit}$ equal to 0.1, 0.3, 0.5,
0.7, 0.8, 0.9 and 0.99 are shown.}
\label{vevol}
\end{figure}

The variation of the polar radius when the rotation velocity varies is illustrated
in the left part of Fig.~\ref{rprpo}. For the 3, 9 and 20 M$_\odot$ at
solar metallicity, the polar radius decreases when the rotation increases. 
The decrease is at most 1.5\%. For the 60 M$_\odot$ model the behaviour is reversed, the polar radius
slightly increases when rotation increases. The maximum increase amounts to 4.5\%.
This difference of behavior for the 60 M$_\odot$ star is likely linked to the fact that in massive stars
radiation pressure plays a more important role.

In the right part of Fig.~\ref{rprpo}, the variation of the luminosity as a function of the rotation
is shown for 60 M$_\odot$ stellar models on the ZAMS at various metallicities.
One sees that the luminosity decreases by at most 5 to 10\% when rotation increases up to the critical limit. This is due to the fact that
a rotating star is, on the average, more extended. As a consequence, the gradients of the temperature and thus the fluxes are smaller.

\section{Initial conditions for reaching the critical limit}

The conditions for reaching the break-up limit have recently been reviewed by \citet{MV}.
We concentrate here on the case of 20 M$_\odot$ stellar models. The cases of the other
initial masses will be discussed in Ekstr\"om et al. (in preparation).

Figure~\ref{vevol} shows the evolution of the surface velocity expressed as a fraction of
the critical velocity during the Main-Sequence phase. The computations were stopped as soon as the star
reaches the critical limit. Quite generally the curves can be decomposed into three parts, whose importance depends on the
initial rotation and metallicity:
\begin{enumerate}
\item At the very beginning, there is a short adjustment period, which lasts for
a few percents of the Main-Sequence lifetime, during which meridional circulation transports angular momentum from
the outer parts of the star to the inner ones, until an asymptotic profile of the angular velocity is reached \citep{Za92}. As a consequence the surface velocity decreases. For instance, a value of 
$\upsilon/\upsilon_{\rm crit}$ on the ZAMS equal to 0.35 passes to a value of 0.30 or slightly lower
after a few 10$^5$ years. For lower initial values of $\upsilon/\upsilon_{\rm crit}$ the decrease is
much weaker, while for higher values, it is higher. This illustrates the dependence of the meridional velocity on $\Omega$.
\item As explained for instance in \citet{MMAA}, after the first adjustment period, a large outer cell of meridional circulation sets in, transporting angular momentum from the inner parts of the star to the outer ones. If this transport is rapid enough and the
stellar winds not too intense, then, $\upsilon/\upsilon_{\rm crit}$
increases. This is the case for all the models with $\Omega/\Omega_{\rm crit} \ge 0.3$
for $Z \ge 0.00001$. For Pop III stars, the increase of $\upsilon/\upsilon_{\rm crit}$ only occurs
for the model with $\Omega/\Omega_{\rm crit} = 0.7$. This is a consequence of the Gratton-\"Opick effect
making the meridional circulation velocity smaller at lower $Z$.
\item At the end of the Main-Sequence phase, the increase of $\upsilon/\upsilon_{\rm crit}$ accelerates
when the star contracts after the blue hook in the HR diagram.
\end{enumerate}
At solar metallicity, the lower initial value of $\Omega/\Omega_{\rm crit}$ for a 20 M$_\odot$ star
to reach the critical limit during the Main- Sequence phase is about 0.8
($\upsilon_{\rm ini}=\sim 400$ km s$^{-1}$). At $Z$ = 0.002 the limiting value
is around 0.7. At $Z$=0.00001, the limiting value is below 0.7, while for Pop III stars, the
limiting value is about 0.99 ! More precise values will be given in Ekstr\"om et al. (in preparation), however based on these results we can already draw the following consequences:
only stars with a sufficiently high initial velocity will reach the critical limit. This
is well in line with the results obtained by \citet{Martayan} showing that the initial velocities
of the Be stars is significantly higher than the initial velocities of the normal B stars.
For 20 M$_\odot$ stars, the range of initial velocities for which the star reaches the critical limit
during the MS phase is larger at lower $Z$ when $Z$ varies between 0.02 and 0.00001.
For pop III 20 M$_\odot$ stellar models, this range is much smaller than at higher metallicities.
For higher initial mass stars, the luminosity may be sufficiently close from the Eddington limit that
the $\Omega\Gamma$--limit can be reached \citep{MMVI}.

The complete coverage of the ranges of initial masses, metallicities and velocities will allow to make predictions for the frequency of Be stars resulting from single star evolution at various metallicities. 

%The size of the convective core depends very weakly on rotation for the 3 M$_\odot$.
%On the other hand, it presents important variation as a function of the metallicity.
%Starting from the Z=0.02 models, one sees that the size of convective core increases when the metallicity %decreases to
%Z=0.002. Less CNO elements are presents, this is compensated by a larger central temperature.
%When the metallicity decreases further, the convective convective core then decreases. Compare the
%Z=0.002 models with the Z=0.00001 models. 

%\begin{figure}%[!t]
%\plottwo{}{.eps}
%\caption{{\it Left:} Location in the HR diagram where the models arrive on the
%ZAMS, for various metallicities and $\Omega/\Omega_{\rm crit}$ ratios. Note that
%$\Omega/\Omega_{\rm crit}$ is crescent redwards. {\it Right:} Same as left, but
%in the log$T_{\rm c}$-log$\rho_{\rm c}$ diagram.
%}
%\label{hr}
%\end{figure}

%\begin{figure}[!t]
%\plotfiddle{teff7.eps}{7cm}{-90}{40}{40}{-160}{250}
%\plotone{B15rc.ps}
%\caption{}
%\label{teff}
%\end{figure}

%\acknowledgements %%% Text of acknowledgements runs on after this command.

%%% THE BIBLIOGRAPHY
%%%
%%% CONSULT SECTION 3 OF "INSTRUCTIONS FOR AUTHORS" FOR HOW TO USE NATBIB.
%%% AUTHORS ARE ENCOURAGED TO USE EITHER THE "THEBIBLIOGRAPY" ENVIRONMENT
%%% BY UNCOMMENTING (DELETING THE "%" SYMBOL) THE COMMANDS BELOW, OR BY
%%% USING THE BIBTEX ENVIRONMENT. TO FIND OUT WHICH IS APPLICABLE TO YOUR
%%% CONTRIBUTION, CONSULT THE VOLUME EDITORS FOR YOUR PROCEEDINGS.
%%%

\end{document}